\documentclass{article}
\usepackage{amsmath,amssymb}
\usepackage[utf8]{inputenc}
\usepackage{graphicx}
\usepackage{bm}
\usepackage{authblk}
\usepackage[numbers]{natbib}
\begin{document}
\title{Stable Calculation of Optical Properties of
  Large Non-Periodic Dissipative Multilayered Systems}
\author[1,2]{Luis Eduardo Puente-D\'\i az}
\author[1,2,3]{Victor  Castillo-Gallardo}
\author[4]{Guillermo P. Ortiz}
\author[5]{Jos\'e Samuel P\'erez-Huerta}
\author[1]{H\'ector P\'erez-Aguilar}
\author[3]{Vivechana Agarwal}
\author[2]{W. Luis Moch\'an}
\affil[1]{Facultad de Ciencias Físico Matem\'aticas,
      Universidad Michoacana de San Nicolás de Hidalgo, Av. Francisco
      J. M\'ugica S/N 58030, Morelia, Mich., M\'exico.}
\affil[2]{Instituto de Ciencias F\'\i sicas, Universidad Nacional Aut\'onoma
  de M\'exico, Av. Universidad S/N, Col. Chamilpa, 62210 Cuernavaca,
  Morelos, M\'exico.}
\affil[3]{Centro de Investigación en Ingenier\'\i a y Ciencias Aplicadas,
  Universidad del Estado de Morelos, Av. Universidad 1001
  Col. Chamilpa, Cuernavaca, Morelos 62209, M\'exico.}
\affil[4]{Departamento de F\'\i sica, Facultad de Ciencias Exactas,
  Naturales y Agrimensura, Universidad Nacional del Nordeste,
  Av. Libertad 5460, 3400, Corrientes, Argentina.}
\affil[5]{Unidad Acad\'emica de Ciencia y Tecnología de la Luz y la
  Materia, Universidad Aut\'onoma de Zacatecas, Carretera
  Zacatecas-Guadalajara km. 6, ejido la Escondida, Campus UAZ Siglo
  XXI, Zacatecas, Zac. 98160, México.}
\maketitle

\begin{abstract}
  The calculation of the transfer matrix for a large non-periodic
  multilayered system may become unstable in the presence of absorption.
  We discuss the origin of this instability and we explore two methods
  to overcome it: the use of a total matrix to solve for all the
  fields at all the interfaces simultaneously and an expansion in the Bloch-like
  modes of a periodic artificially repeated system. We apply both methods to
  obtain the reflectance spectra of multilayered chirped
  structures composed of nanostructured porous silicon (PS). Both methods yield
  reliable and numerically stable results. The former allows an
  analysis of the field within all layers while the latter is much more
  efficient computationally, allowing the design of novel structures
  and the optimization of their parameters. We compare numerical and experimental
  results across a wide spectral range from the infrared to the ultraviolet.
\end{abstract}

\section{Introduction}

Multilayered dielectric structures have been studied extensively in the
visible (Vis) and  near infrared  (NIR) frequency ranges, as in these
ranges these structures might have a small dispersion and dissipation, and
might thus have a high reflectivity compared to that of metallic mirrors
\cite{Fink1998}. This high reflectivity may be achieved over a
wide range of frequencies and angles of incidence and for both TE
and TM polarization in what are known as omnidirectional mirrors (OM's)
\cite{Fink1998,Lekner2000}. The simplest and most common
OM is a structure formed through the periodic repetition of a unit cell formed
from two alternating layers with high and low refractive indices
\cite{Fink1998,Winn98}. There are many examples of these OM's designed
for the Vis
\cite{Weihua2005,Guan2011,Ariza2012,JENA2016} and NIR
\cite{Fink1998,Bruyant2003,Park2003,Valligatla2012} ranges. One common
material for their manufacture is nanostructured porous silicon (PS), obtained from Si
wafers through an electrochemical etching, a
simple synthesis technique that does not require sophisticated
equipment \cite{Escorcia2007,01Ariza_Flores_2011}. This process
allows a control of the
refractive indices of the layers that make up our structure through
the current
density applied during the anodizing process. The thickness of the
layers is controlled through the time during
which this current is applied.
Some works have reported omnidirectional dielectric mirrors
composed of multilayered PS structures in which the index of
refraction varies quasi-continuously according to a given functional
dependence on the depth
\cite{Estevez2008,Estevez2009}. Other have stacked two or more periodic
structures, each composed of pairs of layers with different thicknesses
\cite{Xifre2005,Xifre2009} yielding NIR OM's.  Completely oxidized
{\em chirped} multilayered structures, that is, multilayered structures where the
thickness of successive pairs of layers is gradually increased \cite{Ariza2012},
have also been developed for the Vis region.

The calculation of the optical properties of these systems is usually
carried out through the use of transfer matrices
\cite{Pochi2005,02Ariza_Flores_2011}. Each layer is characterized by a $2\times2$ matrix
that transfers the continuous independent components of the
electromagnetic field from one interface to the next. Multiplying the
matrices of all the layers we obtain a transfer matrix that relates
the fields at the first and last interfaces, where boundary conditions
are applied to obtain the optical coefficients. Unfortunately, in the presence of
dissipation, the simple product of transfer matrices may become
unstable \cite{Perez2004} and the resulting optical properties may be unreliable. This
would be the case for chirped PS OM tuned to the ultraviolet range in
which Si shows a non-negligible dispersion and dissipation. In order
to attack this and similar cases, in this paper we explore two alternative methods
to achieve numerical stability, reliability and computational
efficiency.

The paper is organized as follows. In Section \ref{theory} we develop
two alternative formalisms that allow the calculation of the optical
properties of large multilayered structures even in the case where the
ordinary transfer matrix method fails; in Subsection \ref{ssExt} we
present a formalism based on an extended matrix that allows
calculating the fields at all interfaces, while in Subsection
\ref{ssBloch} we present a formalism based on an expansion on
Bloch-like modes of an artificial periodically repeated structure from
which the actual system is a finite slice. In Section
\ref{Experimental-details} we provide experimental details about our
manufacture of the porous silicon structures with which we test our
formalism and in Section \ref{Results} we present and discuss numerical and
experimental results. Finally, Section \ref{Conclusions} is devoted to
conclusions.

\section{Theory}\label{theory}
\subsection{Transfer Matrix}\label{ssM}
\begin{figure}
  $$
  \includegraphics[width=.7\textwidth]{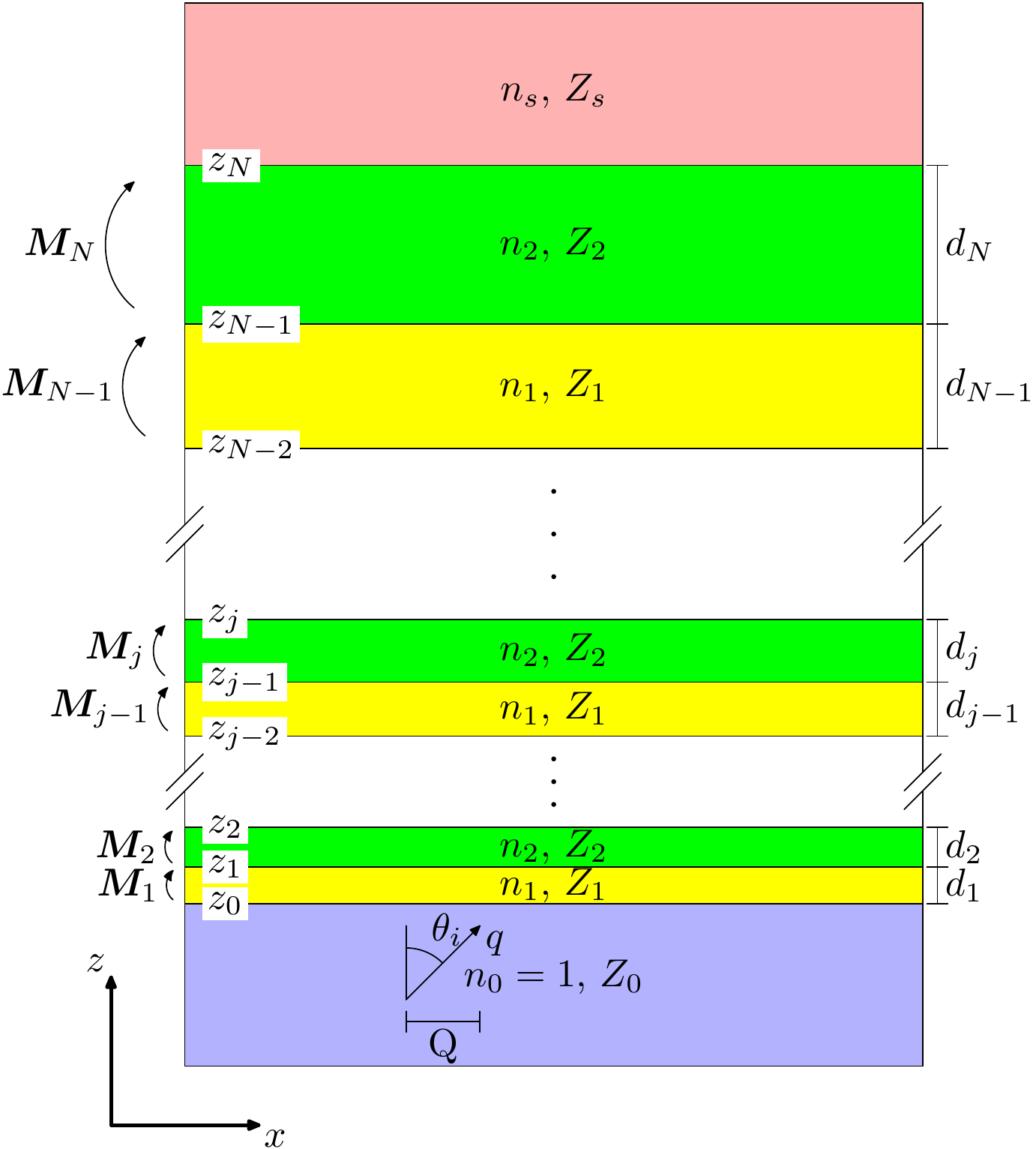}
  $$
  \caption{\label{esquema}Multilayered system composed of $N$ layers numbered
    $j=1\ldots N$ with widths $d_j$ and characterized by a transfer
    matrix $\bm M_j$. Each layer is made of one or another of two
    alternating materials with indices of refraction $n_1$ and $n_2$.
    The system is deposited on a substrate with index of refraction
    $n_s$ and is illuminated from an ambient with index of refraction
    $n_0$ which we take as vacuum ($n_0=1$). All interfaces lie on the
    $xy$ plane, the axis of the structure is along $z$ and we took
    $xz$ as the incidence plane. We indicate the height $z_j$ of each
    interface, the wavevector of the incident field with wavenumber
    $q=\omega/c=2\pi/\lambda$, incidence angle $\theta_i$ and parallel
    projection $Q=q\sin\theta_i$.
    }
\end{figure}

Let us consider a  system  composed of $N$ layers $j=1\ldots N$ of
width $d_j$ and index of refraction $n_j$, as shown in
Fig. \ref{esquema}, with interfaces lying on the $xz$ plane and
stacked along the $z$ direction.
The propagation of an electromagnetic
wave in this system can be described by a $2\times2$ transfer matrix
\begin{equation}\label{Mtot}
\bm M=\bm M_N \bm M_{N-1}\ldots \bm M_2\bm M_1
\end{equation}
that relates the components parallel to the interfaces,  $E_\|$ and $H_\|$,
of the electric and magnetic fields across the structure
\begin{equation}\label{MEH}
\begin{pmatrix}
    E_\|\\H_\|
\end{pmatrix}_{z_N}=
\bm M
\begin{pmatrix}
    E_\|\\H_\|
\end{pmatrix}_{z_0},
\end{equation}
where we designate by $z_{j-1}$ the lower and by $z_j$ the upper interfaces of layer $j$,
$z_0$ corresponds to the interface with the ambient of index of
refraction $n_0$, $z_N$ to the interface with the substrate of
index of refraction $n_s=n_{N+1}$, and $\bm M_j$ relates the fields across
a single layer, from $z_{j-1}$ to $z_j$,
\begin{equation}\label{MEHi}
\begin{pmatrix}
    E_\|\\H_\|
\end{pmatrix}_{z_j}=
\bm M_j
\begin{pmatrix}
    E_\|\\H_\|
\end{pmatrix}_{z_{j-1}},
\end{equation}
and is given by
\begin{equation}\label{Mi}
\bm M_j=
\begin{pmatrix}
\cos k_jd_j & iZ_j\sin k_jd_j \\
iY_j\sin k_jd_j & \cos k_jd_j
\end{pmatrix}
,
\end{equation}
where
\begin{equation}\label{kj}
  k_j=\sqrt{\epsilon_j q^2-Q^2}
\end{equation}
is the $z$ component of the
wavevector for fields that move towards the $z$
direction, $Z_j$ is the corresponding surface impedance,
$Y_j=1/Z_j$ is the surface admittance, $\epsilon_j=n_j^2$ is the
permittivity (for simplicity we assumed
nonmagnetic media with permeability $\mu_j=1$),
$q=\omega/c=2\pi/\lambda$ is
the free-space wavenumber corresponding to the wavelength
$\lambda$ and $\bm Q$ is the projection of the
wavevector onto the interfaces, which is conserved according to
Snell's law and the law of reflection, $Q=n_0 q\sin\theta_i$ with
$\theta_i$ the angle of incidence.
The surface impedances are given by $Z_j=q /k_j$ for TE
polarization and $Z_j=k_j /q\epsilon_j$ for TM
polarization.

By writing the fields at the ambient at $z_0^-$ in terms of an incident and a
reflected wave,
\begin{equation}\label{EHvsr}
  E_\|(z_0)=
  \begin{cases}
    1+r&\text{(TE)}\\
    Z_0(1-r)&\text{(TM)}
  \end{cases},\quad
  H_\|(z_0)=
  \begin{cases}
    Y_0(1-r)&\text{(TE)}\\
    1+r&\text{(TM)}
  \end{cases},\quad
\end{equation}
and writing the fields in the substrate at $z_N^+$ in
terms of a transmitted wave
\begin{equation}\label{EHvst}
  E_\|(z_N)=
  \begin{cases}
    t&\text{(TE)}\\
    Z_st&\text{(TM)}
  \end{cases},\quad
  H_\|(z_N)=
  \begin{cases}
    Y_st&\text{(TE)}\\
    t&\text{(TM)}
  \end{cases},\quad
\end{equation}
assuming an incident wave of unit amplitude, Eq. (\ref{MEH}) becomes a
system of two equations which may be solved for the two unknowns, the
reflection and transmission amplitudes $r$ and $t$.

\subsection{Extended Matrix}\label{ssExt}

The common procedure above is very simple and efficient and works well
for many systems. Nevertheless, in the cases of absorptive layers, for
which $\epsilon_j$ has an imaginary part, and for metallic systems or
right above a resonance for dielectric systems, for which
$\epsilon_j$ may be negative, the wavevector components
$k_j$ may become complex. This may be the case even for transparent
systems in the case where $Q$ is so large that the arguments of the
square roots in Eq. (\ref{kj}) become negative. In this case, the
trigonometric functions in the transfer matrix (Eq. (\ref{Mi})) get an
exponential contribution. Upon the multiplication of many of them to
get the transfer matrix of the whole system (Eq. (\ref{Mtot})), all
the matrix elements would grow exponentially with the size of the
system, yielding extremely ill-conditioned matrices that may become
useless for the accurate computation of optical properties.

We notice that even when there is dissipation,  the
transfer matrix of each layer ought to be unimodular, i.e., $\det \bm
M_j=1$ \cite{Perez2004}. Thus, $\bm M$ should also be unimodular and its two
eigenvalues ought to be mutually inverse. Nevertheless, in the
presence of dissipation, all elements of
the transfer matrix would be large. Thus, there ought to be an exquisite
cancellation of large terms in the determinant to yield the value 1.
Small numerical noise would destroy this cancellation precluding the
accurate calculation of the smallest eigenvalue.

An alternative to the procedure above is to use an extended or complete
$2N\times2N$ matrix \cite{Perez2004} instead of a single $2\times2$ transfer
matrix, in order to solve simultaneously the set of $2N$
equations (\ref{MEHi}) together with Eqs. (\ref{EHvsr}) and
(\ref{EHvst}) for $r$, $t$, and the
fields $E_\|(z_j)$ and $H\|(z_j)$ at all internal interfaces
$j=1\ldots N-1$. Thus we solve an equation of the form
\begin{equation}\label{LFI}
  \bm L\bm F=\bm I,
\end{equation}
where
\begin{equation}\label{F}
\bm{F}=
(r, E_{\| }(z_1), H_{\| }(z_1), \ldots E_{\| }(z_{N-1}), H_{\| }(z_{N-1}),t)^T
\end{equation}
contains information about the field everywhere within the structure,
\begin{equation}\label{I}
\bm I=(I_{1},I_{2},\ldots 0,0)^T
\end{equation}
is the inhomogeneous driving term, with
\begin{equation}\label{Iij}
I_{1}=
\begin{cases}
-m_{11}^{1}-Y_0m_{12}^{1}&\text{(TE)}\\
-Z_0m_{11}^{1}-m_{12}^{1}&\text{(TM)}
\end{cases},\quad
I_{2}=
\begin{cases}
-m_{21}^{1}-Y_0m_{22}^{1}&\text{(TE)}\\
-Z_0m_{21}^{1}-m_{22}^{1}&\text{(TM)}
\end{cases},
\end{equation}
(we denote by the superscript $T$ the transpose of a matrix) and
\begin{equation}\label{L}
\bm L=
\left(
  \begin{array}{cccccccc}
    \bm L_1&-\bm 1_{2\times2}&\bm0_{2\times2}&\bm 0_{2\times2}&\cdots&\cdots&\cdots&\bm 0_{2\times1}\\
    \bm 0_{2\times1}&\bm M_2&-\bm 1_{2\times2}&\bm 0_{2\times2}&\cdots&\cdots&\cdots&\bm 0_{2\times1}\\
    \bm 0_{2\times1}&\bm 0_{2\times2}&\bm M_3&-\bm 1_{2\times2}&\cdots&\cdots&\cdots&\bm 0_{2\times1}\\
    \vdots&\vdots&\vdots&\vdots&\ddots&\ddots&\ddots&\vdots\\
    \bm 0_{2\times1}&\bm 0_{2\times2}&\cdots&\cdots&\cdots&\bm M_{N-1}&-\bm 1_{2\times2}&\bm 0_{2\times1}\\
    \bm 0_{2\times1}&\bm 0_{2\times2}&\cdots&\cdots&\cdots&\cdots&\bm M_N&\bm L_2\\
  \end{array}
\right)%
\end{equation}
is a large sparse matrix coupling the field components among
themselves, which we write in blocks, where $\bm 1_{2\times2}$ is the
unit $2\times2$ matrix, $\bm 0_{2\times1}$ and $\bm 0_{2\times2}$ are
a $2\times1$ and $2\times2$ matrices of zeroes, and we defined
\begin{equation}
  \label{Lij}
  \begin{aligned}
    \bm L_1=&
    \begin{cases}
      (m_{11}^{1}-Y_0m_{12}^{1},m_{21}^{1}-Y_0m_{22}^{1})^T,&\text{(TE)}\\
      (-Z_0m_{11}^{1}+m_{12}^{1},-Z_0m_{21}^{1}+m_{22}^{1})^T,&\text{(TM)}
    \end{cases}\\
    \bm L_2=&
    \begin{cases}
      -(1,Y_s)^T,&\text{(TE)}\\
      -(Z_s,1)^T,&\text{(TM)}
    \end{cases}
  \end{aligned}
\end{equation}
where we denote by $m^j_{kl}$ the $k,l$-th element of the matrix $\bm M_j$.

Many standard methods may be employed to solve Eq. (\ref{LFI}), such
as the Gaussian elimination \cite{Analysis1993,Olschowk1996}, Gauss-Jordan
\cite{Algebra1971,Analysis1993}, Choleski
\cite{Kershaw1978,Analysis1993}, conjugate gradient
\cite{Magnus1952,Analysis1993,Algebra1971}, and generalized minimal
residual methods \cite{Saad1986,Analysis1993}, among others. Some may take
advantage for the sparseness and tridiagonality by blocks of the matrix
of coefficients in
Eq. (\ref{L}), while others may not. In any case, these
methods include pivoting strategies that judiciously choose the
sequence of steps to take in simplifying the system of equations in
order to numerically stabilize the solution procedure. The usual
transfer matrix formalism is equivalent to an immediate elimination of
all the fields $E_\|(z_j)$ and $H_\|(z_j)$, $j=1\ldots N-1$,  which a
priori may not
turn out to be the best strategy with regards to the numerical
stability of the solution. For this reason, we expect that the solution
of Eq. (\ref{LFI}) may be accurately obtained in systems for which
that of Eqs. (\ref{MEH}), (\ref{EHvsr}) and (\ref{EHvst}) may not.

\subsection{Bloch Expansion}\label{ssBloch}

In the previous subsection we presented a method for obtaining the
optical coefficients of a layered structure together with the fields
at all its interfaces, that we expect would be more stable than the
common transfer matrix method of subsection \ref{ssM}. Nevertheless,
it implies a much larger computational load. This may be a bagatelle
for a single calculation, but it may be of importance, for example,
when designing
an optimized structure through a minimization procedure that requires
full spectra to be calculated for all candidate sets of
design parameters. For this reason, in this subsection we develop an
alternative method.

To this end, we take the complete multilayered system of
Fig. \ref{esquema}, and we replicate it
periodically to form an infinite artificial photonic crystal.
We can then use
Bloch's theorem to describe the normal modes of this crystal.
According to Bloch's theorem, the modes of a
periodic system may be written as a superposition of Bloch waves, each
of which acquires a {\em phase factor} as
it propagates from one period to the next. Therefore, each Bloch wave
would obey
\begin{equation}\label{Bloch}
\begin{pmatrix}
E_\|^\pm \\
H_\|^\pm
\end{pmatrix}
_{z_N}=
\bm M
\begin{pmatrix}
E_\|^\pm \\
H_\|^\pm
\end{pmatrix}
_{z_0}
=
e^{\pm iKD}
\begin{pmatrix}
E_\|^\pm \\
H_\|^\pm
\end{pmatrix}
_{z_0},
\end{equation}%
where $D=z_N-z_0$ is the period, which corresponds to the actual
thickness of the multilayered system, and $\pm K$
represents a 1D
Bloch's vector corresponding to a wave that propagates along the $\pm z$
direction \cite{PerezHuerta2018,LuisMochan1987,LuisMochan1988}. Thus,
$\Lambda_\pm=e^{\pm iKD}$ are the eigenvalues of the transfer matrix
$\bm M$ and
$(E_\|^\pm, H_\|^\pm)^T$ are the corresponding eigenvectors. Notice
that we have used the fact that $\det \bm M =1$ exactly, so that the
product of the eigenvalues is $\Lambda_+\Lambda_-=1$, and the
dispersion relation of the Bloch modes may be obtained in principle from
\begin{equation}\label{dispersion}
  \cos KD=\frac{1}{2} \text{tr}\,\bm M,
\end{equation}
where $\text{tr}$ denotes the trace.

\begin{figure}
  $$
  \includegraphics[width=.7\textwidth]{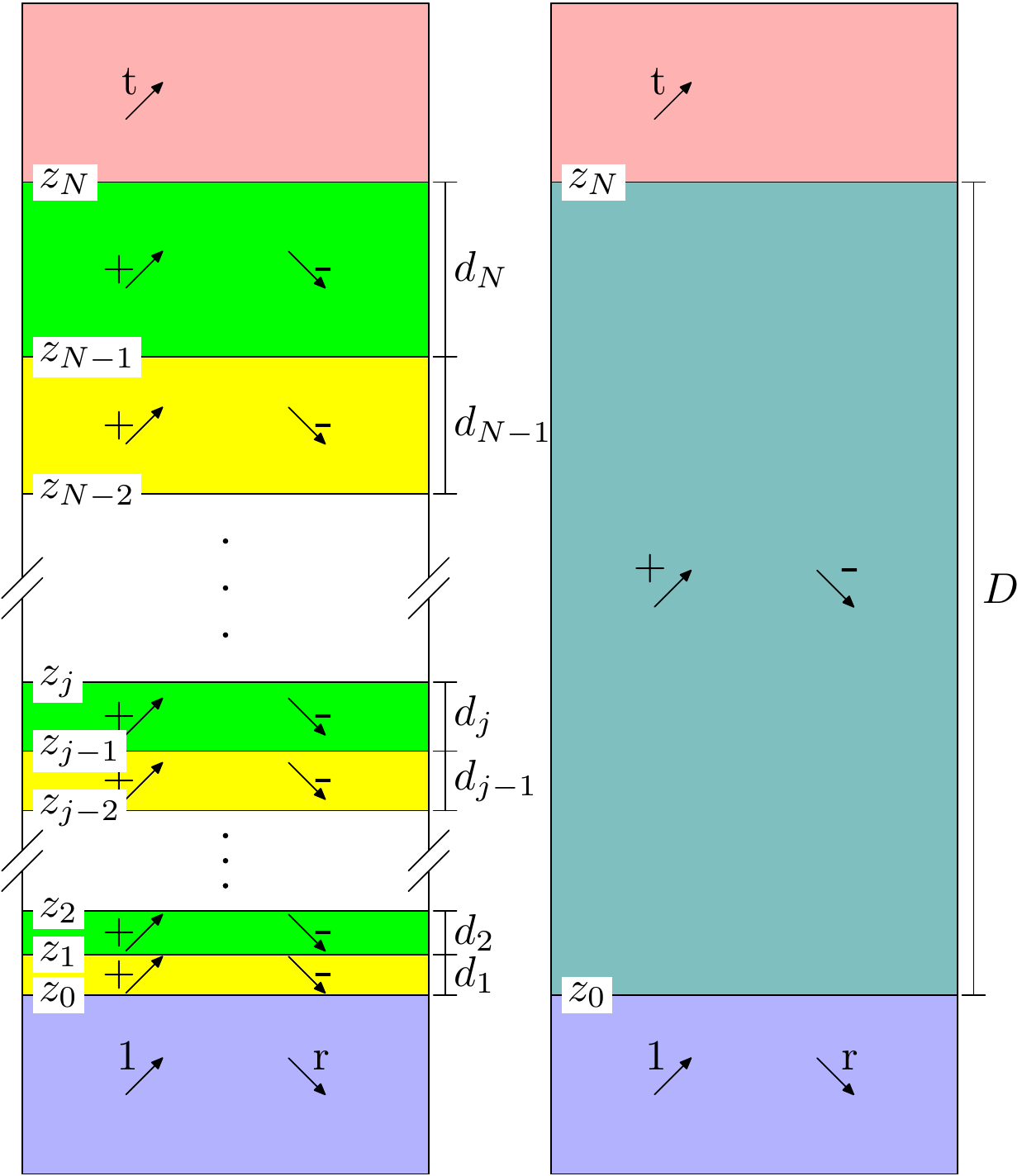}
  $$
  \caption{\label{fBloch}A wave of amplitude 1 is incident from the
    ambient into the
    surface of a multilayered system where it is partially reflected into the
    ambient and transmitted towards the substrate. Within each layer
    there are waves propagating upwards ($+$) and downwards ($-$). The
    multilayered system may be replaced by an effective wide layer
    within which there is one Bloch wave propagating upwards and
    another propagating downwards. The optical coefficients $r$ and
    $t$ and the amplitudes of the Bloch waves may be obtained by
    applying boundary conditions on $E_\|$ and $H_\|$ at $z_0$ and
    $z_N$.}
\end{figure}

Consider now a finite system of width $MD$ made by stacking together
$M$ periods on a substrate. In this case, periodicity would be lost,
and a single Bloch mode would not solve the electromagnetic wave
problem. Nevertheless, the upwards moving Bloch wave would be
reflected downwards at the interface with the substrate, and a
downwards moving Bloch wave would be reflected upwards at the
interface with the ambient. Thus, the optical properties of a finite
system may be obtained by considering a wave incoming from the
ambient, a wave reflected back towards the ambient, a wave transmitted
towards the substrate and two Bloch waves within the multilayered
system, one moving upwards and one moving downwards, as illustrated in
Fig. \ref{fBloch} for the extreme case of only $M=1$ period, which is
the case we analyze below.
The continuity of
$E_\|$ and $H_\|$ would yield two equations at the two interfaces,
with the ambient and with the substrate, from
which we may obtain the four unknowns, namely, $r$, $t$ and the
amplitudes of both Bloch waves. Notice that for an infinite system,
Bloch's vector $K$ should be real, as the Bloch wave would otherwise
diverge either as $z\to\infty$ or $z\to-\infty$. Nevertheless, for a
finite system we may use {\em Bloch-like} modes, for which we allow $K$ to be complex.

In the presence of even a very tiny dissipation, the Bloch-like waves should
decay as they propagate. Thus, we identify the eigenvalue of the
upwards moving mode $\Lambda_+=e^{iKD}$ as the one that obeys
$|\Lambda_+|<1$, $\text{Im}K>0$,
adding a negligible amount of dissipation if necessary to resolve the
apparent ambiguity when $|\Lambda_+|=1$. Similarly, the downwards moving wave has
\begin{equation}\label{Lambda->1}
  |\Lambda_-|>1.
\end{equation}
Notice that we may avoid the numerical instability
issues discussed above if we first identify the eigenvalue for the
downwards wave
\begin{equation}\label{lambda-}
\Lambda _{-}=\frac{1}{2} \text{tr} \bm M \pm i\sqrt{1-(\text{tr}\bm M/2)^2},
\end{equation}
where we chose the sign so as to obey Eq. (\ref{Lambda->1}), and then
obtain the eigenvalue for the upward wave
\begin{equation}\label{lambda+}
  \Lambda _{+}=\frac{1}{\Lambda_-}.
\end{equation}
Following this procedure we avoid the cancellations which amplify the
numerical noise and we obtain eigenvalues that are consistent with the
exact unimodularity of the transfer matrix.

Having obtained the eigenvalues $\Lambda_\pm$ of the transfer matrix,
we may obtain the corresponding eigenvectors $E_\|^\pm$ and $H_\|^\pm$
from Eq. (\ref{Bloch}), and from them, the corresponding surfaces
impedances
\begin{equation}\label{Z+-}
  Z^\pm=-\frac{M_{12}}{M_{11}-\Lambda _{\pm }},
\end{equation}
where $M_{ij}$ ($i,j=1,2$) denote the elements of the
transfer matrix $\bm M$. By writing the fields at $z_0$ and $z_N$ as a superposition of upward and downward propagating (or decaying) fields $E_\|^\pm=Z^\pm H_\|^\pm$, we can relate the fields at $z_N$ to the fields at $z_0$ through a {\em reconstructed} transfer matrix,
\begin{equation}\label{tilMEH}
  \begin{pmatrix}
    E_\|\\H_\|
  \end{pmatrix}_{z_N}=
  \tilde{\bm M}
  \begin{pmatrix}
    E_\|\\H_\|
  \end{pmatrix}_{z_0},
\end{equation}
where
\begin{equation}\label{tilM}
  \tilde{\bm M}=
  \frac{1}{Z^+-Z^-}
  \begin{pmatrix}
    Z^+\exp(iKD)-Z^-\exp(-iKD)& -2iZ^+Z^-\sin KD \\
    2i\sin KD & Z^+\exp(-iKD)-Z^-\exp(iKD)
\end{pmatrix}.
\end{equation}
It can be shown that this matrix complies with unimodularity.

The result above can be readily generalized to a system of $MN$ layers
made up of $M>1$ repetitions of an arbitray structure with $N$
layers. To that end it is only necessary to interpret $\bm M$ in
Eqs. \eqref{dispersion} and \eqref{lambda-} as the transfer matrix of
one period, substitute $z_{N}$ by $z_{MN}$ in
Eq. \eqref{tilMEH} and D by $MD$ in Eq. \eqref{tilM}.

We can finally solve Eqs. (\ref{EHvsr}), (\ref{EHvst}), (\ref{tilMEH})
and (\ref{tilM}) to obtain explicit expressions for the optical
coefficients
\begin{equation}\label{rBloch}
r=\mp \frac{Z_{0}\tilde{M}_{11}+\tilde{M}_{12}-Z
_{0}Z _{s}\tilde{M}_{21}-Z _{s}\tilde{M}_{22}}{Z _{0}
\tilde{M}_{11}-\tilde{M}_{12}-Z _{0}Z _{s}\tilde{M}
_{21}+Z _{s}\tilde{M}_{22}},
\end{equation}
and
\begin{equation}\label{tBloch}
t=\frac{2Z_{\alpha}}{Z _{0}
	\tilde{M}_{11}-\tilde{M}_{12}-Z _{0}Z _{s}\tilde{M}
	_{21}+Z _{s}\tilde{M}_{22}},
\end{equation}
where we choose the upper sign $-$ in Eq. (\ref{rBloch}) and the
subscript $\alpha=s$ in Eq. (\ref{tBloch})
for the case of
TE polarization, while the lower sign $+$ and the subscript $\alpha=0$
correspond to TM polarization. As usual, the
reflectance is given by $R=|r|^2$ and the transmittance by $T=\beta
\left\vert t\right\vert ^{2}$ with $\beta =Z_{0}/Z_{s}$  for
the case of TE polarization and $\beta =Z_{s}/Z_{0}$ for the
case of TM polarization.
Notice that one may factor out and cancel from Eq. (\ref{rBloch}) a
possibly large factor $e^{-iKD}$ and that the dominant term when
$\text{Im}\, KD$ is large is
\begin{equation}\label{rSemi}
  r\approx\pm\frac{Z^+-Z_0}{Z^++Z_0},
\end{equation}
which coincides with the result for a semi-infinitely repeated system, as there
would be a negligible contribution from the Bloch-like wave reflected at
the substrate. In this case, instead of starting the calculation above
from $\bm M$ it may be enough to start from a partial
transfer matrix
\begin{equation}\label{M1tot}
\bm M'=\bm M_{N'} \bm M_{N'-1}\ldots \bm M_2\bm M_1
\end{equation}
with $N'<N$, but large enough so that the interface at $z_{N'}$ is
beyond the reach of the upward-moving Bloch's wave.

\section{Experimental details}\label{Experimental-details}

A photonic structure was synthesized through anodic
etching of a (100) oriented, p-type Boron doped, crystalline Si wafer
with resistivity 0.002-0.005 $\Omega \cdot$cm, under galvanostatic
conditions \cite{Canham1990,Escorcia2007}. The electrochemical
anodizing process was performed at room temperature, with an
electrolyte mixture of aqueous HF (48\% (w/w)) and ethanol (99.9\%
(w/w)) in 1:1 volumetric proportion, respectively. The current density
and the etching duration of each layer was controlled using a
programmable current source.
The current densities were chosen as 2 and 305 mA$/$cm$^{2}$, with
corresponding porosities 41\% and 76\%, respectively. The calibration
curves were acquired through a gravimetric technique as follows: Silicon wafers were used
for synthesizing under similar conditions single layers of porous
silicon, their weights $m_i$ were determined before ($m_1$) and
after ($m_2$) the electrochemical attack, and after
dissolving the already formed porous silicon layer ($m_3$), to
calculate the porosity as $p=(m_1-m_2)/(m_1-m_3)$
\cite{Brugeman2006}. The rate of formation of the nanostructured porous silicon films
was obtained by synthesizing again single layers under similar
conditions and measuring their
thicknesses  through  scanning electron microscopy (SEM).
The absolute reflectivity measurements were carried out with a Perkin
Elmer Lambda 950 UV/Visible spectrophotometer with a variable angle
universal reflectance accessory (URA) for different incident
angles $\theta_i=8^\circ$, $30^\circ$, $45^\circ$ and $60^\circ$ using
non-polarized light. The maximum and minimum values of $\theta_i$ were
constrained by the angular range of the URA.

\section{Results and discussion}\label{Results}

\begin{figure}
\begin{center}
\includegraphics[width=\textwidth]{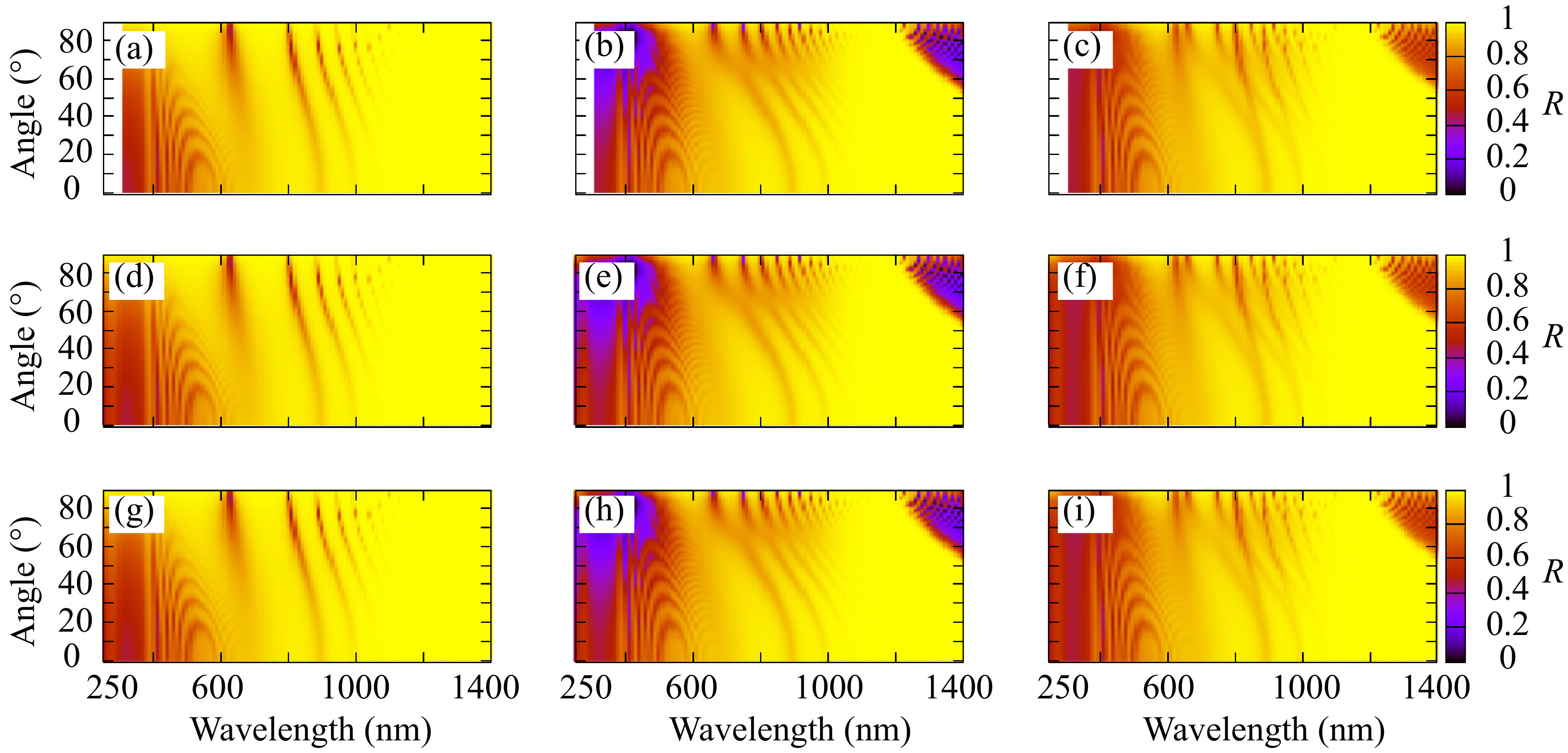}
\end{center}
\caption{\label{Figura4}
  Reflectance spectra of a structure of 101 pairs of PS layers,
  calculated as a function of angle of incidence
  $\theta_i$ and wavelength $\lambda$ using the transfer matrix
  (upper panels), the extended matrix (middle), and the
  Bloch expansion methods (lower) for TE (left), and TM (center)
  polarizations, and for non-polarized light (right). The layers have
  porosities $p_1=41\%$ and $p_2=76\%$ and widths obtained from
  Eqs. (\ref{1/4}) and (\ref{lambdaD}) choosing $\lambda_i=400$ nm,
  $\lambda_f=1400$ nm and $\nu=0.35$.
}
\end{figure}
In Fig. \ref{Figura4} we show the reflectance spectra calculated with
the three methods discussed in Sec. \ref{theory} for a system made up
of $P=101$ pairs of PS layers with alternating porosities $p_1=41\%$
and $p_2=76\%$ respectively on a Si substrate. The thicknesses
$d_{2k-1}$ and $d_{2k}$ ($k=1\ldots P$) were chosen to correspond to
quarter-wave plates,
\begin{equation}\label{1/4}
  k^D_{2k-1}d_{2k-1}= k^D_{2k}d_{2k}=\pi/2,
\end{equation}
as in a {\em Bragg mirror} \cite{Bruyant2003,Xifre2009},
where $k^D_{2k-1}$
and $k^D_{2k}$ were obtained from Eq. (\ref{kj})
evaluated at a given depth dependent {\em design} wavelength \cite{Ariza2012}
\begin{equation}\label{lambdaD}
\lambda^D_k=\lambda _{i}+(\lambda _{f}-\lambda
_{i})\left(\frac{k-1}{P-1}\right)^{\nu }
\end{equation}
with initial wavelength
$\lambda _{i}=400$ nm and final wavelength $\lambda _{f}=1400$ nm.
The value of the exponent $\nu =0.35$ was
chosen by maximizing the calculated reflectance averaged over the wavelengths 250
nm-1400 nm and the angles $0^\circ-90^\circ$,
respectively. To that end, we used the Nelder-Mead
\cite{Nelder1965,Lagarias1998} {\em simplex} method through
the MINUIT package \cite{Minuit2004}. This is a widely used simple but
robust optimization
algorithm. The optimal average reflectance we obtained was 0.91.
Eq. (\ref{lambdaD}) has been shown to yield {\em chirped} multilayered
structures with high reflectance over  a wide frequency range \cite
{Ariza2012}. The refractive indices of the nanostructured PS layers were
obtained for each wavelength using the Bruggeman effective medium
theory \cite{Brugeman2006} and a wavelength dependent Si response
\cite{Palik_1998,02DataBase2015}.
We notice that using the $2\times2$ transfer matrix, the reflectance
spectra could not be calculated for wavelengths $\lambda<310$ nm for
which Si becomes highly dissipative and the double-precision transfer matrix overflowed
numerically (white regions in upper row of
Fig. \ref{Figura4}).

The results of the extended matrix method coincide closely with those
of the standard transfer matrix where the latter converges. Furthermore, it
converges with no problem over all the range explored, down to and beyond
$\lambda=250$ nm. The results of using the Bloch expansion method are
indistinguishable from those of the extended
matrix.
Thus, despite the fact that the standard transfer matrix method is
very useful and commonly used, it fails when the system is highly
dissipative or is made up of a very large number of layers. The
extended matrix and the Bloch expansion methods do not have this
limitation. Moreover, they coincide among themselves and coincide
with the transfer matrix method whenever it converges. Although
numerical stability is obtained when working with the
extended matrix, the computation time it requires is much larger than
that of the $2\times2$ matrices. The time may be somewhat reduced by
reducing the number of unknowns by aggregating the layers in groups
characterized by a single matrix, given by the product of the transfer
matrices of its members, as numerous as possible as long as that the
determinant of the
transfer matrix of the group does not drift away from the nominal value
1. Even applying this grouping separately for the spectral region where Si is
highly dissipative, requiring many small groups, and where it is not,
for which a few large groups suffice, the
computation time required is much larger than that using
the Bloch expansion. Thus, we
conclude that the Bloch expansion provides us with numerical
stability, reliability and computational efficiency. An advantage,
though, of the extended matrix, is that it yields the
field profiles, as illustrated below.

In Fig. \ref{Figura6} we show the squared magnitude of the electric
field as a function of depth for the case of a TE field incident on
the same structure as in Fig. \ref{Figura4} for various angles of
incidence and wavelengths, obtained by using the extended
matrix.
\begin{figure}
  \begin{center}
    \includegraphics[width=\textwidth]{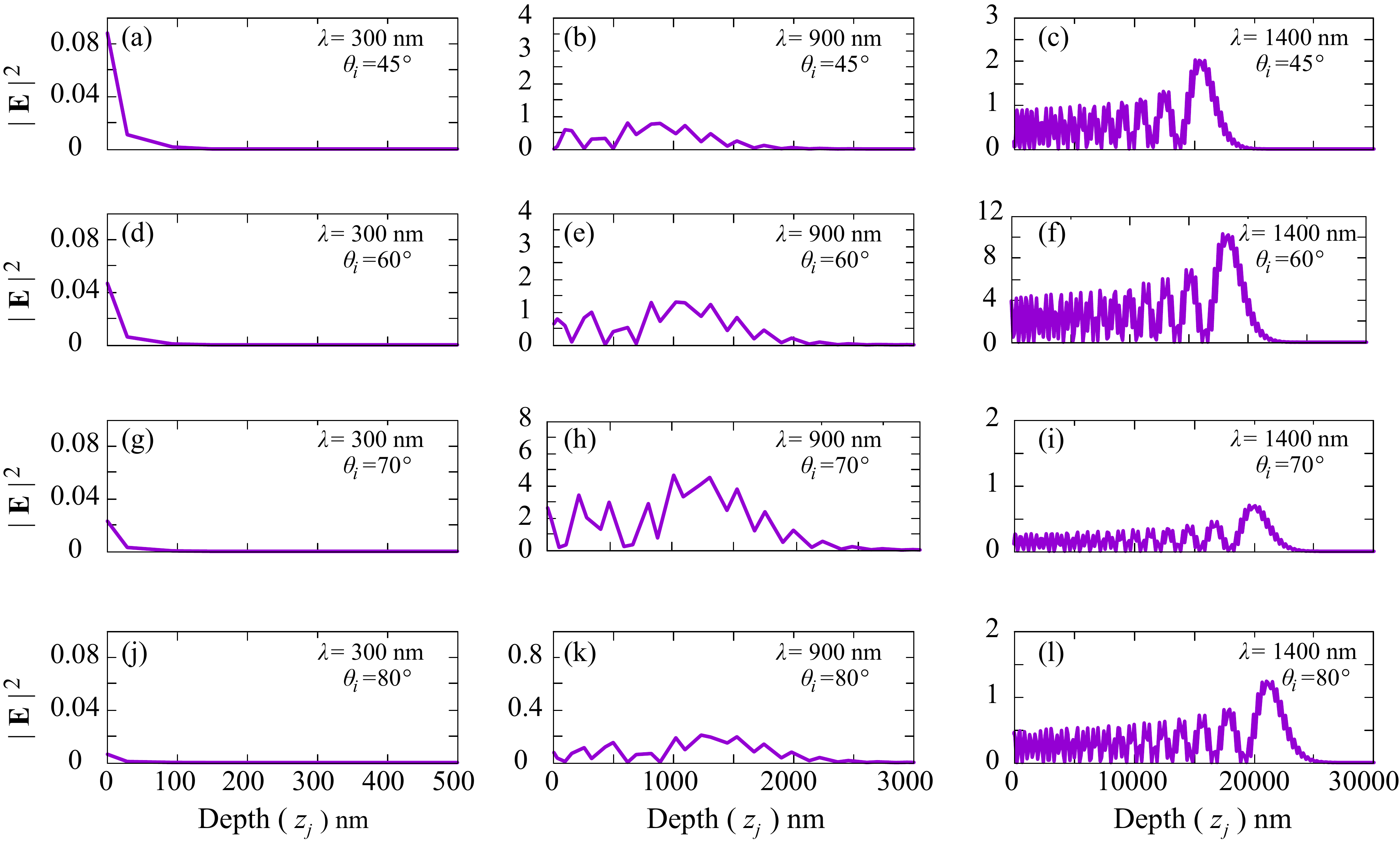}
  \end{center}
  \caption{Squared magnitude of the electric field for TE
    polarization as a function of depth for same system as in
    Fig. \ref{Figura4} for different angles of incidence
    ($\theta_i=45^\circ$, $60^\circ$, $70^\circ$, and $80^\circ$ from
    top to bottom) and several wavelengths ($\lambda=300$  nm, 900 nm,
    and 1400 nm from left to right).
  }
  \label{Figura6}
\end{figure}
We observe that the penetration depth of the electromagnetic
field increases as the wavelength increases. This
is not unexpected, as we designed our structure with thicker layers
deeper inside. The penetration depth also increases as the angle of
incidence increases. Notice the oscillations in the field profile,
more notable for the cases with deeper penetration. There are short
lengthscale oscillations corresponding to the texture of the
structure, and longer lengthscale oscillations due to the interference
between multiply reflected waves from the region where propagation is
forbidden due to Bragg reflections and from the front surface of the
structure. These long-scale oscillations are responsible for the
oscillations visible in the reflectance spectra for long wavelengths
in Fig. \ref{Figura4}.

We remark that while the total thicknesses of this structure is
$D\approx32$ $\mu$m,  the penetration depth
turns out to be no larger than 25 $\mu$m, covering just 83 of the 101
periods, in the case $\lambda =1400$ nm, $\theta_i=80^\circ$.

In the corresponding case but for TM polarization, the
field penetrates a much larger distance,
\begin{figure}
\begin{center}
\includegraphics[width=\textwidth]{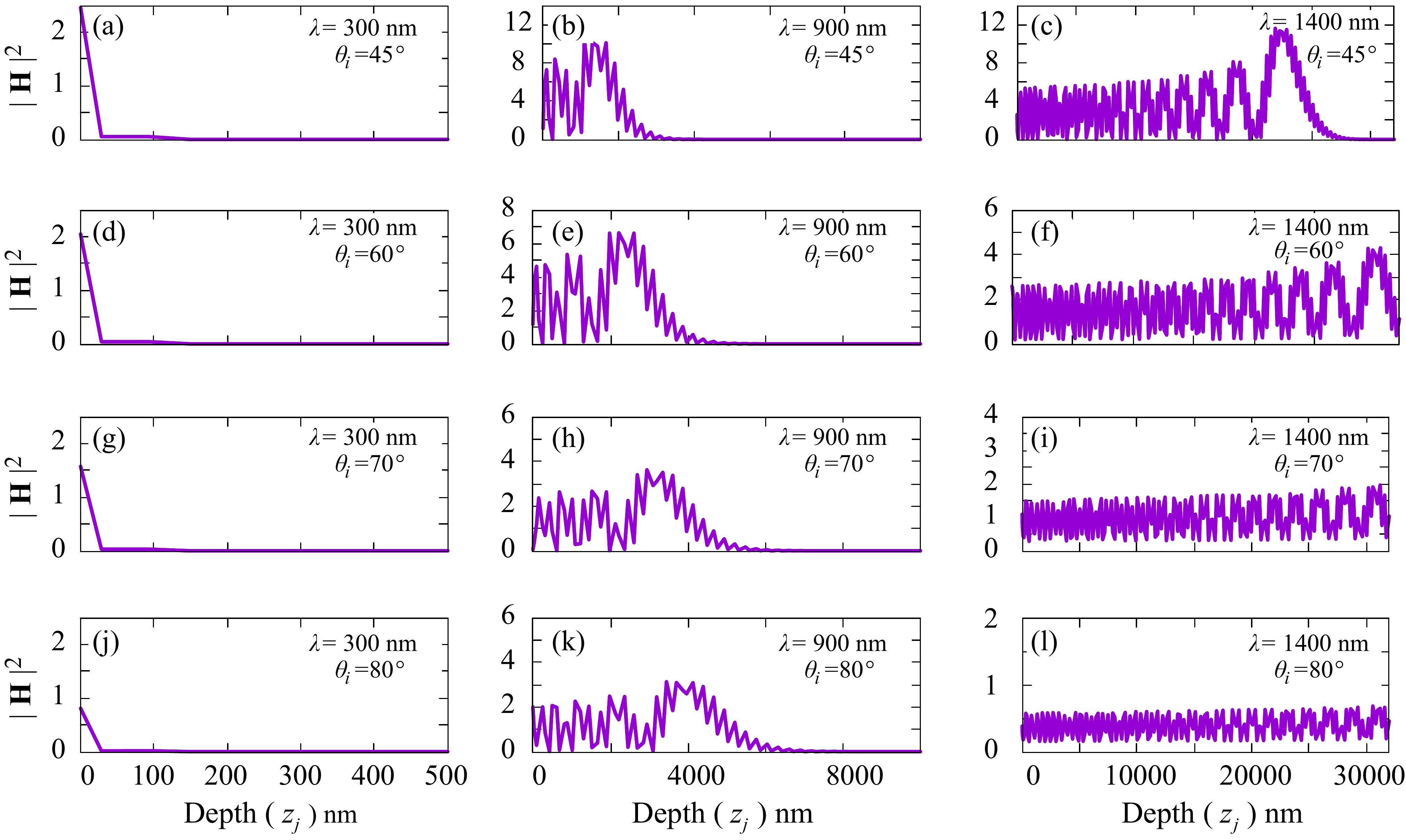}
\end{center}
\caption{\label{Figura7} Squared magnitude of the magnetic field as a
  function of depth, as in Fig. \ref{Figura6}, but for TM polarization.}
\end{figure}
as shown Fig. \ref{Figura7} for the same
structure as in Fig. \ref{Figura6}. In this case, the field
already penetrates more than 25 $\mu$m for $\lambda =1400$ nm and
$\theta_i=45^\circ$, while for larger angles it penetrates the entire
structure.

The results above suggest that for some combinations of polarization,
wavelength and angle of incidence, smaller structures may produce the
same results than the full structures discussed previously.
\begin{figure}
\begin{center}
\includegraphics[width=\textwidth]{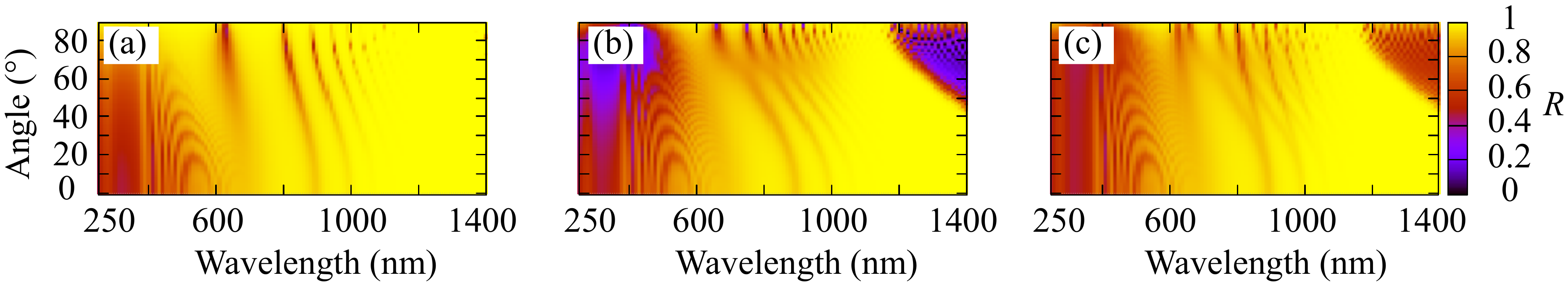}
\end{center}
\caption{\label{Figura10}
Reflectance spectra as a function of wavelength and angle of incidence
obtained for TE (left), TM (center) polarizations, and for
non-polarized light (right), using the Bloch expansion for a structure
as in Fig. \ref{Figura4} but with only the first 83 periods.
}
\end{figure}
In Fig. \ref{Figura10} we show the reflectance spectra
calculated for TE and TM polarizations, and for non-polarized light
using the Bloch expansion for a multilayered system
consisting of the first 83 periods of the system corresponding to
Fig. \ref{Figura4} with 101 periods. According to
Fig. \ref{Figura6}, this system is wider than
the penetration depth for TE polarization and $\theta_i<80^\circ$ and
we can observe the expected correspondence between
Figs. \ref{Figura10}a-c with Figs. \ref{Figura4}g-i.
We have verified this agreement quantitatively. The agreement for TE
polarization is better than for TM,
given the smaller penetration depth. Even though we considered here
narrower systems, the usual transfer
matrix method failed in the UV, while the extended
matrix and the Bloch expansion methods succeeded and were consistent.

\begin{figure}
  \begin{center}
    \includegraphics[width=\textwidth]{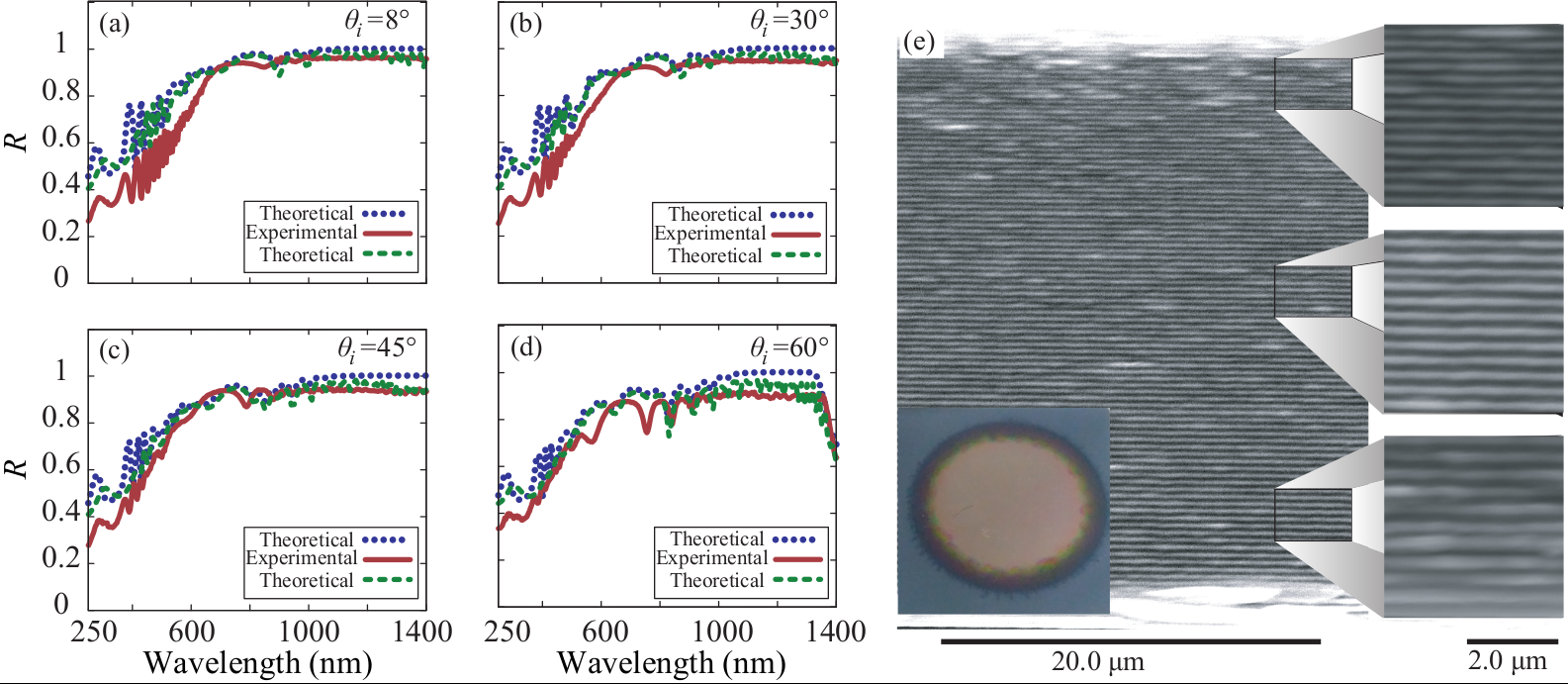}
  \end{center}
  \caption{\label{Exp1}
    Measured (solid red)
    and calculated (dotted blue) reflectivity
    spectra of the sample described in the text as a function of wavelength at four
    different angles (a) $8^\circ$, (b) $30^\circ$, (c)
    $45^\circ$ and (d) $60^\circ$. We also include a modified
    calculation that accounts for some roughness and confinement
    effects (green dashed, see text).
    (e) Micrograph of the
    cross-section of the 101-period structure. The inset shows a
    top view photograph.
  }
\end{figure}

In Fig. \ref{Exp1} we show the experimental
reflectance for non-polarized incident light as a
function of wavelength for various angles of incidence. The fabricated
sample corresponds to the structure presented in Fig. \ref{Figura4}, and
consists of 101 pairs of layers with target
porosities 41\% and 76\%, as described in
Sec. \ref{Experimental-details}, and with target widths obtained from
Eqs. (\ref{1/4}) and (\ref{lambdaD}) with $\lambda_i=400$~nm,
$\lambda_f=1400$~nm and $\nu=0.35$.
SEM images of the synthesized structure are also shown in the figure,
displaying the gradual increase  in the thickness of the layers with
increasing depth.
For comparison, Fig. \ref{Exp1} also shows theoretical results obtained as in
Fig. \ref{Figura4}. Notice that
we could calculate $R$ for the lowest wavelengths only through our proposed formalisms.
The calculated and measured spectra have similar features, though the
experimental reflectance is lower, more so at larger angles, and their
differences are also larger at shorter wavelengths, where the
theoretical reflectance shows larger oscillations.
The differences between the experimental and calculated spectra could
be partially due to the scattering of light at the actual interfaces,
which naturally have some roughness \cite{GuillermoOrtiz2020}.
They may also be due to confinement induced
changes in the dielectric function of the Si phase of porous silicon,
as it has been argued \cite{theiB_1994,THEB1997} that in the blue
spectral range the imaginary part of the
response of the solid phase of heavily p-type doped porous silicon is
significantly larger than that of bulk
silicon, and that its interband transitions become
broadened and red-shifted. Thus, in Fig. \ref{Exp1} we also show
theoretical results obtained as in Fig. \ref{Figura4} but
incorporating some effects of roughness through a macroscopic interface
transfer matrices \cite{GuillermoOrtiz2020} and convoluting the
dielectric function of the Si phase \cite{Palik_1998,02DataBase2015} with a
Gaussian weight in order to red-shift and broaden its spectral
features. The parameters of the modified theory are the
roughness height, and the red-shift and broadening of the Si
response.
Thus, we fitted a shift of 0.032 $\mu$m
and a spectral broadening of 0.022 $\mu$m, of the order of those
discussed in
Refs. \cite{theiB_1994,THEB1997}, and a depth dependent roughness
amplitude that increases grows from 0.5 to 2.6 nm from ambient towards
substrate in
proportion to the widths of each layer. The modified theory yields a
lower reflectance overall, diminished oscillations for short
wavelengths and the decay of the reflectance of large wavelengths and
large angles, in good agreement with experiment.
In Fig. \ref{Exp1} we also illustrate the profile of the structure
through a SEM image. The gradual increase of the thickness of the
layers with increasing depth is visible and is consistent with our
design given by Eq. (\ref{lambdaD}).


\section{Conclusions}\label{Conclusions}
We have shown that the usual transfer matrix approach to the calculation
of optical properties of multilayered systems may fail for large
systems in the presence of absorption.
Nevertheless, numerical stability may be achieved by using 2 alternative
methods: using an extended
or complete transfer matrix and using a method based on the excitation of
Bloch-like modes. We applied these methods to the calculation of the
reflectance spectrum of a wide spectrum omnidirectional mirror
consisting of a large multilayered
chirped structure made of porous silicon. Both proved to be precise
and stable. The extended matrix is
considerably slower than the
expansion in Bloch modes, but it yields more information and allows an
analysis of the fields dependence on depth, which we applied to study the wavelength and
angle dependent penetration depth, which in turn allowed us to replace the
original structures by a thinner one that yielded the desired optical
properties within a large range of wavelengths and angles of
incidence.  On the other hand, the expansion in Bloch-like modes
allows
fast, stable and accurate calculations, providing thus an ideal method
for the design and optimization of large multilayered structures which
could be made of periodic repetitions of a single unit or completely
aperiodic as in our chirped example.
We compared our calculated reflectance spectra to experimental
results and obtained good agreement even for the UV region where there
is relatively strong absorption. The agreement could be improved
by considering effects such as the interfacial roughness and the
modification of the response of Si within the confined pore walls,
effects that can readily be incorporated into our formalism.
Thus, we conclude that our approach is useful for the computation,
design and analysis of the optical properties of very large
multilayered systems.

\section*{Acknowledgments}
This work was supported by DGAPA-UNAM under grant IN111119 and by
CONACyT under grant A1S-30393. LEPD and VCG
acknowledge a scholarship from CONACyT. GPO acknowledges the support of
ANPCyT-FONCyT through grant PICT-0696-2013 and SGCyT-UNNE trough
grants PI-F008-2014 and PI-18F008. HPA also express his gratitude to the
{\em Coordinación de la Investigación Científica de la Universidad Michoacana
de San Nicolás de Hidalgo}. VCG acknowledges useful discussions
with A. David Ariza-Flores.

\bibliographystyle{unsrt}
\bibliography{Bibliografia}

\begin{thebibliography}{10}

\bibitem{Fink1998}
Yoel Fink, Joshua~N. Winn, Shanhui Fan, Chiping Chen, Jurgen Michel, John~D.
  Joannopoulos, and Edwin~L. Thomas.
\newblock A dielectric omnidirectional reflector.
\newblock {\em Science}, 282(5394):1679--1682, 1998.

\bibitem{Lekner2000}
John Lekner.
\newblock Omnidirectional reflection by multilayer dielectric mirrors.
\newblock {\em J. Opt. A: Pure Appl. Opt.}, 2(5):349--352, 2000.

\bibitem{Winn98}
Joshua~N. Winn, Yoel Fink, Shanhui Fan, and J.~D. Joannopoulos.
\newblock Omnidirectional reflection from a one-dimensional photonic crystal.
\newblock {\em Opt. Lett.}, 23(20):1573--1575, 1998.

\bibitem{Weihua2005}
Lin Weihua, Wang~Guo Ping, and Zhang Suhuai.
\newblock Design and fabrication of omnidirectional reflectors in the visible
  range.
\newblock {\em J. Mod. Opt.}, 52(8):1155--1160, 2005.

\bibitem{Guan2011}
Guan Huihuan, Han Peide, Yang Yanqing, Li~Yuping, Zhang Xue, and Zhang Wenting.
\newblock Omni-directional mirror for visible light based on one-dimensional
  photonic crystal.
\newblock {\em Chin. Opt. Lett.}, 9(7):071603--071603, 2011.

\bibitem{Ariza2012}
A.~David Ariza-Flores, L.~M. Gaggero-Sager, and V.~Agarwal.
\newblock White metal-like omnidirectional mirror from porous silicon
  dielectric multilayers.
\newblock {\em Appl. Phys. Lett.}, 101(3):031119, 2012.

\bibitem{JENA2016}
S.~Jena, R.B. Tokas, P.~Sarkar, J.S. Misal, S.~Maidul Haque, K.D. Rao,
  S.~Thakur, and N.K. Sahoo.
\newblock Omnidirectional photonic band gap in magnetron sputtered tio2/sio2
  one dimensional photonic crystal.
\newblock {\em Thin Solid Films}, 599:138--144, 2016.

\bibitem{Bruyant2003}
Bruyant A., G.~L{\'e}rondel, P.~J. Reece, and M.~Gal.
\newblock All-silicon omnidirectional mirrors based on one-dimensional photonic
  crystals.
\newblock {\em Appl. Phys. Lett.}, 82(19):3227--3229, 2003.

\bibitem{Park2003}
Yeonsang Park, Young-Geun Roh, Chi-O Cho, Heonsu Jeon, Min~Gyu Sung, and J.~C.
  Woo.
\newblock Gaas-based near-infrared omnidirectional reflector.
\newblock {\em Applied Physics Letters}, 82(17):2770--2772, 2003.

\bibitem{Valligatla2012}
Sreeramulu Valligatla, Alessandro Chiasera, Stefano Varas, Nicola Bazzanella,
  D.~Narayana Rao, Giancarlo~C. Righini, and Maurizio Ferrari.
\newblock High quality factor 1-d er3$+$-activated dielectric microcavity
  fabricated by rf-sputtering.
\newblock {\em Opt. Express}, 20(19):21214--21222, 2012.

\bibitem{Escorcia2007}
J.~Escorcia and V.~Agarwal.
\newblock Effect of duty cycle and frequency on the morphology of porous
  silicon formed by alternating square pulse anodic etching.
\newblock {\em Phys. Status Solidi (c)}, 4(6):2039--2043, 2007.

\bibitem{01Ariza_Flores_2011}
A.~David Ariza-Flores, L.~M. Gaggero-Sager, and V.~Agarwal.
\newblock Effect of interface gradient on the optical properties of
  multilayered porous silicon photonic structures.
\newblock {\em Journal of Physics D: Applied Physics}, 44(15):155102, 2011.

\bibitem{Estevez2008}
J.~O. Estevez, J.~Arriaga, M{\'e}ndez~Blas A, and Agarwal V.
\newblock Omnidirectional photonic bandgaps in porous silicon based mirrors
  with a gaussian profile refractive index.
\newblock {\em Appl. Phys. Lett.}, 93(19):191915, 2008.

\bibitem{Estevez2009}
J.~O. Estevez, J.~Arriaga, A.~M{\'e}ndez Blas, and V.~Agarwal.
\newblock Enlargement of omnidirectional photonic bandgap in porous silicon
  dielectric mirrors with a gaussian profile refractive index.
\newblock {\em Appl. Phys. Lett.}, 94(6):061914, 2009.

\bibitem{Xifre2005}
E.~Xifr{\'e}-P{\'e}rez, L.~F. Marsal, J.~Pallarès, and J.~Ferr{\'e}-Borrull.
\newblock Porous silicon mirrors with enlarged omnidirectional band gap.
\newblock {\em Journal of Applied Physics}, 97(6):064503, 2005.

\bibitem{Xifre2009}
E.~Xifr{\'e}-P{\'e}rez, L.~F. Marsal, J.~Ferr{\'e}-Borrull, and
  J.~Pallar{\`e}s.
\newblock Low refractive index contrast porous silicon omnidirectional
  reflectors.
\newblock {\em Applied Physics B}, 95(1):169--172, 2009.

\bibitem{Pochi2005}
P.~Yeh.
\newblock {\em Optical Waves in Layered Media}.
\newblock Wiley, USA, 2nd edition, 2005.

\bibitem{02Ariza_Flores_2011}
David. Ariza-Flores, L.~M. Gaggero-Sager, and V.~Agarwal.
\newblock Omnidirectional photonic bangap in dielectric mirrors: a comparative
  study.
\newblock {\em Journal of Physics D: Applied Physics}, 45(1):015102, 2011.

\bibitem{Perez2004}
R.~Perez-Alvarez and F.~Garcia-Molina.
\newblock {\em Transfer Matrix, Green Functions and Related Techniques}.
\newblock Castello de la Plana: Publicacions de la Universitat Jaume I, 2004.

\bibitem{Analysis1993}
J.~Stoer and R.~Bulirsch.
\newblock {\em Introduction to Numerical Analysis}.
\newblock Springer, NewYork, 3rd edition, 1993.

\bibitem{Olschowk1996}
Markus Olschowka and Arnold Neumaier.
\newblock A new pivoting strategy for gaussian elimination.
\newblock {\em Linear Algebra and its Applications}, 240:131--151, 1996.

\bibitem{Algebra1971}
J.~H. Wilkinson and C.~Reinsch.
\newblock {\em Handbook for Automatic Computation: Volume II: Linear Algebra}.
\newblock Springer-Verlag Berlin Heidelberg, NewYork, 1st edition, 1971.

\bibitem{Kershaw1978}
David~S. Kershaw.
\newblock The incomplete cholesky—conjugate gradient method for the iterative
  solution of systems of linear equations.
\newblock {\em Journal of Computational Physics}, 26:43--65, 1978.

\bibitem{Magnus1952}
M.~R. Hestenes and E.~Stiefel.
\newblock Methods of conjugate gradients for solving linear systems.
\newblock {\em J. of Res. Nat. Bur. Standards}, 49:409--436, 1952.

\bibitem{Saad1986}
Youcef Saad and Martin~H. Schultz.
\newblock Gmres: A generalized minimal residual algorithm for solving
  nonsymmetric linear systems.
\newblock {\em SIAM J. Sci. and Stat. Comput.}, 7(3):856--869, 1986.

\bibitem{PerezHuerta2018}
J.~S. P{\'e}rez-Huerta, D.~Ariza-Flores, R.~Castro-Garc{\'\i}a, W.~L.
  Moch{\'a}n, G.~P. Ortiz, and V.~Agarwal.
\newblock Reflectivity of {1D} photonic crystals: A comparison of computational
  schemes with experimental results.
\newblock {\em Int. J. Mod. Phys. B}, 32(11):1850136, 2018.

\bibitem{LuisMochan1987}
W.~Luis Moch{\'a}n, Marcelo del Castillo-Mussot, and Rub{\'e}n~G. Barrera.
\newblock Effect of plasma waves on the optical properties of metal-insulator
  superlattices.
\newblock {\em Phys. Rev. B}, 35(3):1088--1098, 1987.

\bibitem{LuisMochan1988}
W.~Luis Moch{\'a}n and Marcelo del Castillo-Mussot.
\newblock Optics of multilayered conducting systems: Normal modes of periodic
  superlattices.
\newblock {\em Phys. Rev. B}, 37(12):6763--6771, 1988.

\bibitem{Canham1990}
Canham L.~T.
\newblock Silicon quantum wire array fabrication by electrochemical and
  chemical dissolution of wafers.
\newblock {\em Appl. Phys. Lett.}, 57(10):1046--1048, 1990.

\bibitem{Brugeman2006}
Pap~Andrea Edit, Kord{\'a}s Kriszti{\'a}n, V{\"a}h{\"a}kangas Jouko,
  Uusim{\"a}ki Antti, Lepp{\" a}vuori Seppo, Pilon Laurent, and Szatm{\'a}ri
  S{\'a}ndor.
\newblock Optical properties of porous silicon. part iii: Comparison of
  experimental and theoretical results.
\newblock {\em Optical Materials}, 28(5):506--513, 2006.

\bibitem{Nelder1965}
J.~A. Nelder and R.~Mead.
\newblock A simplex method for function minimization.
\newblock {\em The Computer Journal}, 7:308--313, 1965.

\bibitem{Lagarias1998}
Jeffrey~C. Lagarias, James~A. Reeds, Margaret~H. Wright, and Paul~E. Wright.
\newblock Convergence properties of the nelder--mead simplex method in low
  dimensions.
\newblock {\em SIAM J. Optim.}, 9(1):112--147, 1998.

\bibitem{Minuit2004}
F.~James and M.~Winkler.
\newblock {\em Minuit user’s guide}.
\newblock CERN, Geneva, 2004.

\bibitem{Palik_1998}
E.D. Palik.
\newblock {\em Handbook of Optical Constants of Solids}.
\newblock Elsevier, 1998.

\bibitem{02DataBase2015}
Carsten Schinke, P.~Christian Peest, Jan Schmidt, Rolf Brendel, Karsten Bothe,
  Malte~R. Vogt, Ingo Kr{\"o}ger, Stefan Winter, Alfred Schirmacher, Siew Lim,
  Hieu~T. Nguyen, and Daniel MacDonald.
\newblock Uncertainty analysis for the coefficient of band-to-band absorption
  of crystalline silicon.
\newblock {\em AIP Advances}, 5(6):067168, 2015.

\bibitem{GuillermoOrtiz2020}
Guillermo~P. Ortiz, J~Victor~J. Toranzos, Leandro~A. Missoni, Mar{\'\i}a~L.
  Mart{\'\i}nez-Ricci, and W.~Luis Moch\'an.
\newblock Rough {1D} photonic crystals: a transfer matrix approach.
\newblock Submitted to Journal of Optics. arxiv:2004.00185 [physics.optics].

\bibitem{theiB_1994}
W.~Thei{\ss}, R.~Arens-Fischer, M.~Arntzen, M.G. Berger, S.~Frohnhoff,
  S.~Hilbrich, and Wernke M.
\newblock Probing optical transitions in porous silicon by reflectance
  spectroscopy in the near infrared, visible and {UV}.
\newblock {\em MRS Proceedings}, 358:435, 1994.

\bibitem{THEB1997}
W.~Thei{\ss}.
\newblock Optical properties of porous silicon.
\newblock {\em Surface Science Reports}, 29(3):91--192, 1997.

\end{thebibliography}

\end{document}